\documentclass[prd,onecolumn,showkeys,
preprint,
nofootinbib,
 amsmath,amssymb,
 aps,
]{revtex4-2}
\usepackage{yfonts}
\usepackage{graphicx}
\usepackage{dcolumn}
\usepackage{bm}
\usepackage{nicefrac}
\usepackage{color}
\usepackage{xcolor}
\definecolor{darkblue}{RGB}{0,0,196}
\definecolor{darkgreen}{RGB}{0,120,0}
\usepackage[colorlinks=true,linktocpage=true,linkcolor=darkblue,citecolor=red,urlcolor=darkblue]{hyperref}
\usepackage{hyperref}
\usepackage{appendix}
\usepackage{eqnarray}
\usepackage{subcaption}
\usepackage{array}

\usepackage{fix-cm}

\newcommand{\bea}{\begin{eqnarray}}
\newcommand{\eea}{\end{eqnarray}}
\newcommand{\bel}[1]{\begin{eqnarray}\label{#1}}
\newcommand{\eel}{\end{eqnarray}}
\newcommand{\beq}{\begin{equation}}
\newcommand{\eeq}{\end{equation}}

\def\LB{\left(}
\def\RB{\right)}
\def\LSB{\left[}
\def\RSB{\right]}
\def\LAB{\langle}
\def\RAB{\rangle}

\newcommand{\nn}{\nonumber}

\newcommand{\p}{\partial}
\newcommand{\dd}{\mathrm{d}}

\newcommand{\dotj}{\dot{J}}
\newcommand{\ddj}{\ddot{J}}

\newcommand{\f}[2]{\frac{#1}{#2}}
\newcommand{\onehalf}{{\nicefrac{1}{2}}} 
 
\newcommand{\threefourths}{{\nicefrac{3}{4}}} 

\def\spin{\,\textgoth{s:}}

\begin{document}

\title{Hybrid framework of Fermi-Dirac spin hydrodynamics}

\author{Zbigniew Drogosz}
\email{zbigniew.drogosz@alumni.uj.edu.pl}
\affiliation{Institute of Theoretical Physics, Jagiellonian University, PL-30-348 Krak\'ow, Poland}

\date{\today}

\begin{abstract}
We outline the hybrid framework of spin hydrodynamics, combining classical kinetic theory with the Israel-Stewart method of introducing dissipation.  
We obtain the local equilibrium expressions for the baryon current, the energy-momentum tensor, and the spin tensor of particles with spin 1/2 following the Fermi-Dirac statistics and compare them with the previously derived versions where the Boltzmann approximation was used. The expressions in the two cases have the same form, but the coefficients are governed by different functions. The relative differences between the tensor coefficients in the Fermi-Dirac and Boltzmann cases are found to grow exponentially with the baryon chemical potential.
In the proposed formalism, nonequilibrium processes are studied 
including mathematically possible dissipative corrections. Standard conservation laws are applied, and the condition of positive entropy production allows for transfer between the spin and orbital parts of angular momentum.

\end{abstract}


\maketitle

\section{Introduction}

Experimental results indicating a nonzero spin polarization of particles produced in heavy-ion collisions 
\cite{STAR:2017ckg, STAR:2018gyt, STAR:2019erd, ALICE:2019aid}
have generated a growing interest in observables related to spin degrees of freedom 
\cite{Niida:2024ntm},
motivating at the same time attempts at 
expanding the
theoretical framework of relativistic hydrodynamics \cite{Florkowski:2010zz, Florkowski:2017olj}
to include a description of spin, thus giving rise to the field of spin hydrodynamics. This challenge of developing a new theoretical formalism has been pursued along several paths based on different assumptions:
(a) The final particle spin polarization is determined on the basis of gradients of hydrodynamic fields on the freezeout hypersurface. The vorticity $\varpi_{\mu \nu} =- \f{1}{2} \left(\p_\mu\beta_\nu - \p_\nu \beta_\mu \right)$ and the thermal shear  $\xi_{\mu \nu} =- \f{1}{2} \left(\p_\mu\beta_\nu + \p_\nu \beta_\mu \right)$, where $\beta^\mu = \f{u^\mu}{T}$, are considered to be of particular importance in this case, see Refs.~\cite{Becattini:2009wh,  Becattini:2011zz, Becattini:2013fla, Palermo:2024tza, Becattini:2021iol}. (b) The spin hydrodynamics equations are derived from quantum or classical kinetic theory of particles with spin $\onehalf$ \cite{Florkowski:2017ruc, Florkowski:2017dyn, Bhadury:2020puc, Bhadury:2020cop, Bhadury:2022ulr, Singh:2022ltu, Weickgenannt:2019dks, Weickgenannt:2021cuo, Weickgenannt:2020aaf, Wagner:2022amr, Weickgenannt:2022zxs, Weickgenannt:2023nge, Wagner:2024fhf, Hu:2021pwh, Li:2020eon,  Shi:2020htn}. (c) Following the method of Israel and Stewart \cite{Israel:1979wp}, the laws of conservation and entropy production are applied to general mathematically allowed forms of the energy-momentum and spin tensors \cite{Hattori:2019lfp, Fukushima:2020ucl, Daher:2022xon, Daher:2022wzf, Sarwar:2022yzs, Wang:2021ngp, Biswas:2022bht, Biswas:2023qsw, Xie:2023gbo, Daher:2024ixz, Xie:2023gbo, Ren:2024pur, Daher:2024bah, Gallegos:2021bzp, Hongo:2021ona, Kumar:2023ojl, She:2021lhe}. (d) Lagrangian effective field theory techniques are used in the context of a spin-polarizable medium \cite{Montenegro:2017rbu,Montenegro:2018bcf,Montenegro:2020paq}.

We believe that it is useful and important to reexamine the assumptions and establish relations between the formalisms. Any clarification and synthesis thus achieved may facilitate further progress and, perhaps, eventually lead to removing the present discrepancies.
Hence the recent proposal of a hybrid framework that combines the use of kinetic theory for the perfect fluid description and the Israel-Stewart method for the inclusion of dissipation \cite{Hont2025gen, Drogosz2024hybrid, Florkowski2025spinh}. Among the attractive features of this approach are the simple definition of local thermodynamic equilibrium of particles with spin (conservation of the spin part of the total angular momentum), 
consistent and independent power counting in
the expansion in the spin polarization tensor $\omega^{\mu \nu}$
and in the gradients (the latter appear in dissipative terms, while the second order expansion in the former is necessary already on the perfect-fluid level to obtain nontrivial thermodynamic relations with spin), as well as 
the addition of dissipative effects without resorting 
to the complicated formalism of
nonlocal collisions \cite{Weickgenannt:2019dks, Weickgenannt:2021cuo, Weickgenannt:2020aaf, Wagner:2022amr, Weickgenannt:2022zxs}.

In this contribution, we summarize the hydrodynamic tensors appearing in the hybrid framework, expand the analysis of the more realistic case of the Fermi-Dirac statistics and compare the results with the previously derived Boltzmann approximation.

\section{Results}

\subsection{Tensors of perfect spin hydrodynamics}

We follow the prescription in Ref.~\cite{Drogosz2024hybrid} to derive the tensors of perfect spin hydrodynamics via the kinetic theory. We focus in particular on the case of the Fermi-Dirac statistics.
In the kinetic theory with classical treatment of spin, one introduces the internal angular momentum \cite{Mathisson:1937zz,2010GReGr..42.1011M}
\begin{equation}
s^{\alpha \beta} = \f{1}{m} \epsilon^{\alpha \beta \gamma \delta} p_\gamma s_\delta,
\end{equation}
where the spin 4-vector $s^\mu$ is orthogonal to the 4-momentum $s \cdot p = 0$, $s^\alpha = \f{1}{2m} \epsilon^{\alpha \beta \gamma \delta} p_\beta s_{\gamma \delta}$.
In the particle rest frame, $p^\mu = (m,0,0,0)$, $s^\alpha = (0,\mathbf{s}_*)$, $|\bf{s}_* | = \spin$, and $\spin^2 = \onehalf \left( 1+ \onehalf  \right) = \threefourths$.
The local equilibrium distribution functions for particles ($+$) and for antiparticles ($-$) have the Fermi-Dirac form
\begin{equation}\label{f1}
f^{\pm_{\rm }}_{\rm eq}(x,p,s) = \LSB \exp \LB \mp \xi(x) + p \cdot \beta(x)  -  \frac{1}{2} \, \omega(x) : s \RB +1 \RSB^{-1} = \f{1}{e^{y^\pm} + 1},
\end{equation}
with
\begin{equation}\label{f2}
y^\pm = \mp \xi(x) + p \cdot \beta(x) 
- \frac{1}{2} \omega(x) : s.
\end{equation}
The colon denotes contraction over both indices, $\omega : s \equiv \omega_{\mu \nu} s^{\mu \nu}$. 
The spin polarization tensor is parametrized with the help of 4-vectors $k^\mu$ and $t^\mu$,
\begin{equation}
\omega_{\alpha \beta} = k_\alpha u_\beta - k_\beta u_\alpha + t_{\alpha \beta},
\end{equation}
with
$k \cdot u = \omega \cdot u = 0, t_{\alpha \beta} = \epsilon_{\alpha \beta \gamma \delta} u^\gamma \omega^\delta$, $t^\mu = t^{\mu \nu} k_\nu = \epsilon^{\mu \nu \alpha \beta} k_\nu u_\alpha \omega_\beta.$
{\everymath{\displaystyle}
\begin{table}[t]
\begin{center}
\caption{The baryon current}\label{tab:N}
\begin{tabular}{|ccc|}
\hline
\multicolumn{3}{|c|}{$\vphantom{\bigg|}N_{\rm eq}^\mu = ( n_0+ n_2^\omega +  n_2^k )u^{\mu}+n_{t}t^{\mu}$}                            \\ \hline
\multicolumn{1}{|c|}{Coefficient} & \multicolumn{1}{c|}{Fermi-Dirac} & Boltzmann \\ \hline 
\multicolumn{1}{|c|}{$\vphantom{\bigg|}n_0$} & \multicolumn{1}{c|}{$\f{m^3}{\pi^2} J_{21}^\mp$} & $\f{2\sinh{\xi}}{\pi^{2}}z^{2}T^{3}K_{2}(z)$ \\ \hline
\multicolumn{1}{|c|}{$\vphantom{\bigg|}n_2^\omega$} & \multicolumn{1}{c|}{$ - \omega^2 \f{\spin^2 m^3}{18 \pi^2} (\ddj_{21}^\mp + 2 \ddj_{23}^\mp )$} & $- \omega^2 \f{\spin^{2}\sinh{\xi}}{3\pi^{2}} z T^{3} \LSB z K_{2}(z) + 2 K_{3}(z) \RSB$ \\ \hline
\multicolumn{1}{|c|}{$\vphantom{\bigg|}n_2^k$} & \multicolumn{1}{c|}{$- k^2 \f{\spin^2 m^3}{9\pi^2} \ddj_{41}^\mp$} & $-k^2\f{2 \spin^{2} \sinh{\xi}}{3\pi^{2}} z T^{3}K_{3}(z)$ \\ \hline
\multicolumn{1}{|c|}{$\vphantom{\bigg|}n_t$} & \multicolumn{1}{c|}{$- \f{\spin^2 m^3}{9\pi^2} \ddj_{41}^\mp$} & $-\frac{2\spin^{2}\sinh{\xi}}{3\pi^2}  z T^{3} K_{3}(z)$ \\ \hline
\end{tabular}
\end{center}
\end{table}
}
{\everymath{\displaystyle}
\begin{table}[t]
\caption{The energy-momentum tensor}\label{tab:T}
\begin{tabular}{|ccc|}
\hline
\multicolumn{3}{|c|}{$\vphantom{\bigg|}T^{\mu\nu}_{\rm eq} = (\varepsilon_0 + \varepsilon_2^\omega + \varepsilon_2^k)u^\mu u^\nu - (P_0 +P_2^\omega + P_2^k) \Delta^{\mu \nu} + P_t (t^\mu u^\nu + t^\nu u^\mu)+ P_{k\omega} (k^\mu k^\nu + \omega^\mu \omega ^\nu) $}                            \\ \hline
\multicolumn{1}{|c|}{Coefficient} & \multicolumn{1}{c|}{Fermi-Dirac} & Boltzmann \\ \hline
\multicolumn{1}{|c|}{$\vphantom{\bigg|}\varepsilon_0$} & \multicolumn{1}{c|}{$\f{m^4}{\pi^2 } J_{22}^\pm $} & $\f{2 \cosh \xi}{\pi^2} z^2 T^4 [z K_3(z) - K_2(z)]$ \\ \hline
\multicolumn{1}{|c|}{$\vphantom{\bigg|}\varepsilon_2^\omega$} & \multicolumn{1}{c|}{$- \omega^2 \f{\spin^2 m^4}{18\pi^2}(\ddj_{22}^\pm + 2 \ddj_{24}^\pm)$} & $- \omega^2\f{\spin^2 \cosh \xi}{3 \pi^2} z T^4 [zK_2(z) + (z^2 + 10) K_3(z)] $ \\ \hline
\multicolumn{1}{|c|}{$\vphantom{\bigg|}\varepsilon_2^k$} & \multicolumn{1}{c|}{$- k^2 \f{\spin^2 m^4}{9\pi^2} \ddj_{42}^\pm$} & $-k^2 \f{2 \spin^2 \cosh \xi}{3\pi^2}zT^4[zK_2(z) + 5K_3(z)]$ \\ \hline
\multicolumn{1}{|c|}{$\vphantom{\bigg|}P_0$} & \multicolumn{1}{c|}{$\f{m^4}{3\pi^2} J_{40}^\pm$} & $\f{2 \cosh \xi}{\pi^2}z^2 T^4 K_2(z)$ \\ \hline
\multicolumn{1}{|c|}{$\vphantom{\bigg|}P_2^\omega$} & \multicolumn{1}{c|}{$- \omega^2 \f{\spin^2 m^4}{90 \pi^2} (\ddj_{40}^\pm + 4 \ddj_{42}^\pm)$} & $-\omega^2 \f{\spin^2 \cosh \xi}{3 \pi^2} zT^4 [zK_2(z) + 4K_3(z)] $ \\ \hline
\multicolumn{1}{|c|}{$\vphantom{\bigg|}P_2^k$} & \multicolumn{1}{c|}{$- k^2 \f{2\spin^2 m^4}{45 \pi^2} \ddj_{60}^\pm$} & $- k^2 \f{4 \spin^2 \cosh \xi}{3 \pi^2}zT^4 K_3(z)$ \\ \hline
\multicolumn{1}{|c|}{$\vphantom{\bigg|}P_{k \omega}$} & \multicolumn{1}{c|}{$-\f{\spin^2 m^4}{45 \pi^2} \ddj_{60}^\pm $} & $-\f{2\spin^2 \cosh \xi}{3 \pi^2}zT^4 K_3(z)$ \\ \hline
\multicolumn{1}{|c|}{$\vphantom{\bigg|}P_{t}$} & \multicolumn{1}{c|}{$-\f{\spin^2m^4}{9\pi^2}\ddj_{42}^\pm$} & $-\f{2 \spin^2 \cosh \xi}{3 \pi^2} zT^4 [zK_2(z) + 5K_3(z)]$ \\ \hline
\end{tabular}
\end{table}
}

In local equilibrium, the baryon current, the energy-momentum tensor, and the spin tensor can be written as
\begin{equation}
N^\mu_{\rm eq} = \int \dd P \,\dd S \, p^\mu \, \left[f_{\rm eq}^+(x,p,s)-f_{\rm eq}^-(x,p,s) \right],
\end{equation}
\begin{equation}
T^{\mu \nu}_{\rm eq} = \int \dd P \,\dd S \, p^\mu p^\nu \, \left[f_{\rm eq}^+(x,p,s) + f_{\rm eq}^-(x,p,s) \right],
\end{equation}
\begin{equation}
\hspace{-0.5cm}S^{\lambda, \mu\nu}_{\rm eq} = \int \!\dd P \, \dd S \, \, p^\lambda \, s^{\mu \nu} 
\left[f_{\rm eq}^+(x,p,s)+ f_{\rm eq}^-(x,p,s) \right],
\end{equation}
with the integration measures
\begin{equation}
\dd P = \f{\dd^3 p}{(2\pi)^3 E_p}, \quad
\dd S = \f{m}{\pi \spin} \dd s \delta (s \cdot s + \spin^2) \delta (p \cdot s).
\end{equation}

For small magnitudes of $\omega^{\mu \nu}$, the Fermi-Dirac distribution function (\ref{f1}) can be expanded around $y_s \equiv - \frac{1}{2} \omega : s= 0$
\begin{equation}
f_{\rm eq}^\pm(x,p,s)=\f{1}{e^{y^\pm_0+y_s}+1} = \f{1}{e^{y^\pm_0}+1}-\f{e^{y^\pm_0}}{(e^{y^\pm_0}+1)^2} \, y_s +\f{e^{y^\pm_0} (e^{y^\pm_0}-1)}{2 \, (e^{y^\pm_0}+1)^3} \, y_s^2 \, + \cdots ,
\end{equation}
where $y^\pm_0 \equiv \mp \xi(x) + p \cdot \beta(x)$. Thus,
\begin{align}\begin{split}
f_{\rm eq}^\pm &=  f^\pm_0 - f^\pm_0 (1-f^\pm_0) y_s +  \frac{1}{2}  f^\pm_0 (1-f^\pm_0) (1 - 2 f^\pm_0) y_s^2 + \cdots \\ &\equiv
f^\pm_0 + \frac{1}{2} \, f^\pm_1  \, \omega : s 
+ \frac{1}{8} \,  f^\pm_2  \, (\omega : s)^2  + \cdots
\end{split}\end{align}
Expansion up to the quadratic order in $\omega^{\mu \nu}$ in the case of $N^\mu$ and $T^{\mu \nu}$ (the linear terms vanish), and up to the linear order in $\omega^{\mu \nu}$ in the case of $S^{\lambda, \mu \nu}$ yields
\begin{align}\begin{split}\label{eq:n}
N_{\rm eq}^\mu &=\int \dd P \,\dd S \, p^\mu \, \left[f_{\rm eq}^+(x,p,s)-f_{\rm eq}^-(x,p,s) \right] = 2(Z^{+\mu}_0- Z^{-\mu}_0) \\ &+  \spin^2\f{\omega:\omega}{6} (Z^{+\mu}_2- Z^{-\mu}_2) + \f{\spin^2}{3m^2} (   
Z^{+\mu\alpha\beta}_2 -   Z^{-\mu\alpha\beta}_2 )
\omega^\gamma_{\phantom{\gamma}\alpha} \omega_{\beta \gamma} + ...
\end{split}\end{align}
\begin{align}\begin{split}\label{eq:t}
T_{\rm eq}^{\mu \nu} &= \int \dd P \,\dd S \, p^\mu p^\nu \, \left[f_{\rm eq}^+(x,p,s) + f_{\rm eq}^-(x,p,s) \right] = 
2 \LB Z^{+\mu \nu}_{0}\!+\!Z^{-\mu \nu}_{0} \RB \\ 
&+ \spin^2 \,\f{\omega:\omega}{6} 
\LB Z^{+\mu \nu}_{2}\!+\!Z_{2}^{-\mu \nu} \RB 
\!+\! \f{\spin^2}{3m^2} \,
\LB Z^{+\mu \nu \alpha \beta}_{2}\!+\!Z^{-\mu \nu \alpha \beta}_{2} \RB\omega^\gamma_{\phantom{\gamma}\alpha} \omega_{\beta \gamma} + ...
\end{split}\end{align}
\begin{align}\begin{split}\label{eq:s}
S_{\rm eq}^{\lambda, \mu\nu} &= \int \!\dd P \, \dd S \, \, p^\lambda \, s^{\mu \nu} 
\left[f_{\rm eq}^+(x,p,s)+ f_{\rm eq}^-(x,p,s) \right] 
= \f{2 \spin^2 }{3} \,  \omega^{\mu \nu} \,(Z^{+\lambda}_1 +   Z^{-\lambda}_1) \\ &+ \f{2 \spin^2 }{3m^2} \LSB  \, ( Z^{+ \lambda \alpha \mu }_{1} + Z^{- \lambda \alpha \mu }_{1} )\omega^{\nu}_{\,\,\, \alpha}  -  ( Z^{+ \lambda \alpha \nu }_{1} + Z^{- \lambda \alpha \nu }_{1} )\omega^{\mu}_{\,\,\, \alpha} \RSB + ...
\end{split}\end{align}
with the tensors $Z_n$ defined as
\begin{equation}\label{eq:z}
Z^{\pm \alpha \beta ...}_{n} \equiv \int \dd P p^{\alpha} p^{\beta}... f^{\pm}_n.
\end{equation}
In the Boltzmann approximation, the local distribution function is simpler, $f_{\rm eq}^\pm = \exp(y^\pm)$ (compare Eqs.~(\ref{f1})~and~(\ref{f2})), and the tensors $Z_n$ simplify to the form
\begin{equation}\label{intBoltz}
Z^{\pm \alpha\beta \ldots}_{n} \rightarrow e^{\pm \xi} Z^{\alpha\beta \ldots} \equiv e^{\pm \xi} \int \dd P p^\alpha p^\beta \cdots e^{-\beta \cdot p}, \quad n=0,1,2,\cdots
\end{equation}
{\everymath{\displaystyle}
\begin{table}[t!]
\begin{center}
\caption{The spin tensor}\label{tab:S}
\begin{tabular}{|ccc|}
\hline
\multicolumn{3}{|c|}{$\begin{aligned}\vphantom{\Bigg|}S_{\rm eq}^{\lambda, \mu \nu} &= A_1 u^\lambda \omega^{\mu \nu} + \f{A_2}{2} u^\lambda (u^\mu k ^\nu - u^\nu k^\mu) + \f{A_3}{2} t^{\lambda \mu \nu}\\ \vphantom{\bigg|}&=u^\lambda\big[A_3(k^\mu u^\nu - k^\nu u^\mu) + A_1 t^{\mu\nu}\big] + \f{A_3}{2}\big(t^{\lambda\mu} u^\nu - t^{\lambda \nu} u^\mu + \Delta^{\lambda \mu} k^\nu - \Delta^{\lambda \nu}k^\mu \big)\end{aligned}$} \\ \hline
\multicolumn{1}{|c|}{Coefficient} & \multicolumn{1}{c|}{Fermi-Dirac} & Boltzmann \\ \hline 
\multicolumn{1}{|c|}{$\vphantom{\bigg|}A_1$} & \multicolumn{1}{c|}{$\f{\spin^{2} m^3}{9 \pi^2} ( \dotj_{21}^\pm + 2 \dotj_{23}^\pm)$} & $\f{2\spin^{2}\cosh{\xi}}{3\pi^{2}}zT^{3} \LSB zK_{2}(z)+2K_{3}(z)\RSB$ \\ \hline
\multicolumn{1}{|c|}{$\vphantom{\bigg|}A_2$} & \multicolumn{1}{c|}{$\f{2\spin^{2} m^3}{3 \pi^2} (2\dotj_{23}^\pm - \dotj_{21}^\pm)$} & $\f{4\spin^{2}\cosh{\xi}}{3\pi^{2}}z^2 T^{3}K_{4}(z)$ \\ \hline
\multicolumn{1}{|c|}{$\vphantom{\bigg|}A_3$} & \multicolumn{1}{c|}{$-\f{2 \spin^{2} m^3}{9 \pi^2} \dotj_{41}^\pm$} & $-\f{4\spin^{2}\cosh{\xi}}{3\pi^{2}}z T^{3}K_{3}(z)$ \\ \hline
\end{tabular}
\end{center}
\end{table}
}
\begin{figure}[ht!]
\includegraphics[width= 0.99 \linewidth]{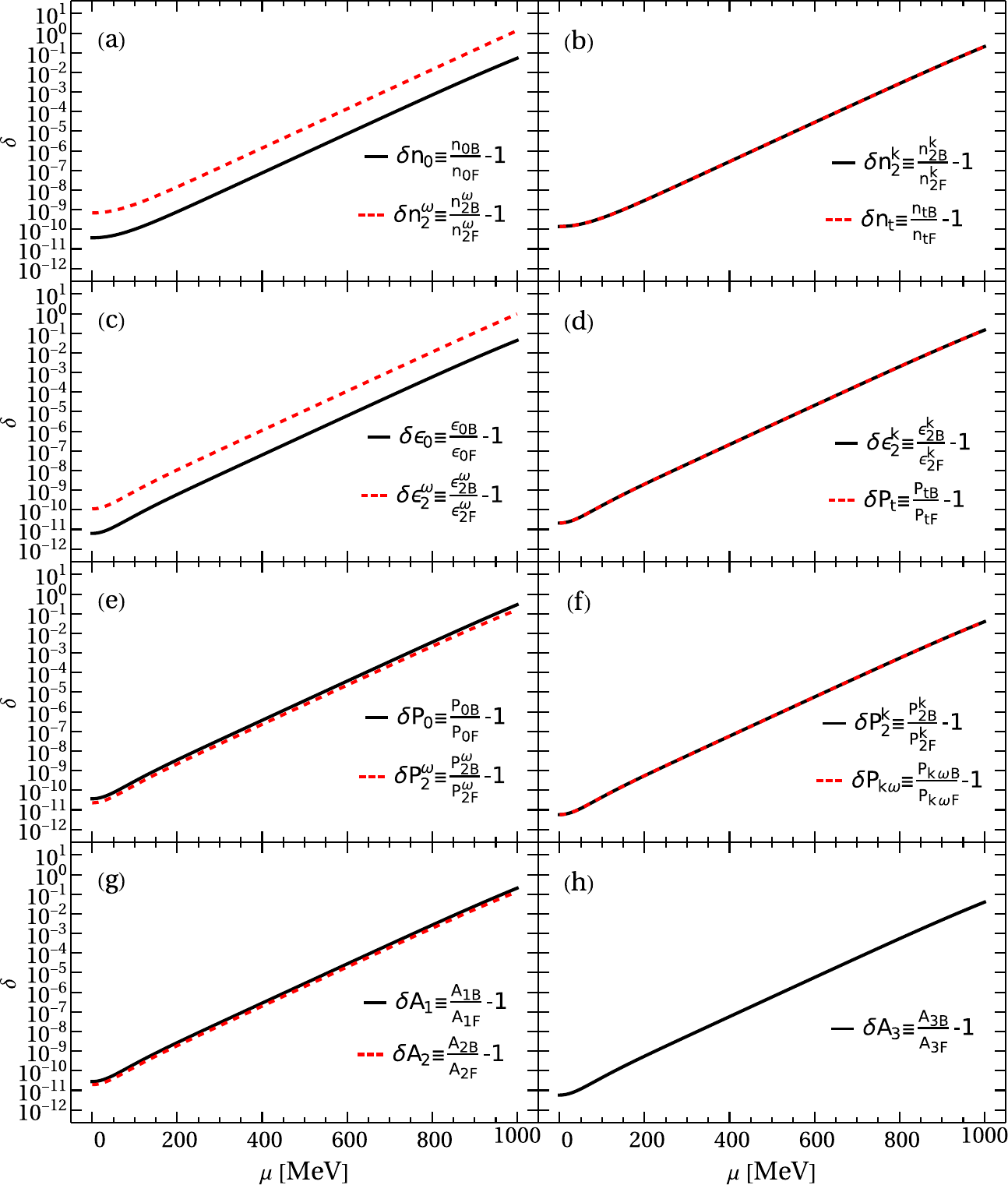}
\caption{Relative differences between the Boltzmann and the Fermi-Dirac case of the coefficients of the baryon current $N_{\rm eq}^\mu$ (a)–(b), the energy-momentum tensor $T_{\rm eq}^{\mu \nu}$ (c)–(f) and the spin tensor $S_{\rm eq}^{\lambda, \mu \nu}$ (g)–(h). Plots for particle mass $m = 1000$ MeV, temperature $T = 100$ MeV, and baryon chemical potential $\mu$ ranging from $0$ to $1000$ MeV.}\label{fig:coeff}
\end{figure}
\noindent
The tensors $Z_0$, $Z_1$ and $Z_2$ with up to four indices were expressed in terms of the function
\begin{equation}
J_{mn}(\xi,z) = \int_0^\infty \f{\sinh^m y \cosh^n y}{\exp(-\xi + z \cosh y) + 1} \dd y, \quad m,n = 0,1,2,\dots
\end{equation}
and its derivatives
in Appendix A of Ref.~\cite{Drogosz2024hybrid}, although subsequently only the Boltzmann approximation was considered. Here we address the Fermi-Dirac case.
We denote the first and the second derivative of the function $J_{mn}$ with respect to $\xi$ with $\dotj_{mn}$ and $\ddj_{mn}$, respectively
\begin{align}\begin{split}
\dotj_{mn}(\xi,z) &\equiv \f{\p}{\p\xi} J_{mn}(\xi,z) = \int_0^\infty \f{\sinh^m y \cosh^n y}{2 + 2\cosh(\xi - z \cosh y)} \dd y,\\
\ddj_{mn}(\xi,z) &\equiv \f{\p^2}{\p\xi^2} J_{mn}(\xi,z) = \int_0^\infty \f{\sinh^m y \cosh^n y \sinh(\xi-z \cosh y)}{2 \big(1 + \cosh(\xi - z \cosh y) \big)^2} \dd y.
\end{split}\end{align}
For brevity, we introduce the following notation using the plus and minus superscript
\begin{align}\begin{split}
J_{mn}^+ \equiv J_{mn}(\xi,z), \quad J_{mn}^- \equiv J_{mn}(-\xi,z),\\
J_{mn}^\pm \equiv J_{mn}^+ + J_{mn}^-, \quad J_{mn}^\mp \equiv J_{mn}^+ - J_{mn}^-
\end{split}\end{align}
and the same for $\dotj$ and $\ddj$.

We gather the expressions derived in Appendix A of Ref.~\cite{Drogosz2024hybrid}, insert them into~(\ref{eq:n})–(\ref{eq:s}), and perform contractions and further simplifications, also noting the usefulness of hyperbolic identities in expressing the functions $J_{mn}$, $\dotj_{mn}$ and $\ddj_{mn}$ by functions with different indices, as needed to convert the expressions into their most compact forms. Finally, we can write the three tensors as
\begin{align}
N_{\rm eq}^\mu &= ( n_0+ n_2^\omega +  n_2^k )u^{\mu}+n_{t}t^{\mu},\\
T^{\mu\nu}_{\rm eq} &= (\varepsilon_0 + \varepsilon_2^\omega+ \varepsilon_2^k)u^\mu u^\nu - (P_0 +P_2^\omega + P_2^k) \Delta^{\mu \nu}\\
&+ P_t (t^\mu u^\nu + t^\nu u^\mu) + P_{k\omega} (k^\mu k^\nu + \omega^\mu \omega ^\nu),\nn\\ \label{spintensor}
S_{\rm eq}^{\lambda, \mu \nu} &= u^\lambda \LSB A \LB k^\mu u^\nu - k^\nu u^\mu \RB + A_1 t^{\mu\nu} \RSB  + \frac{A}{2} \LB t^{\lambda \mu} u^\nu - t^{\lambda \nu} u^\mu + \Delta^{\lambda \mu} k^\nu - \Delta^{\lambda \nu} k^\mu \RB,
\end{align}
with $\Delta^{\mu \nu} \equiv g^{\mu \nu} - u^\mu u^\nu$.
Their form is the same in the Fermi-Dirac and the Boltzmann case, but the functional dependence of the coefficients on $m$, $T$ and $\mu$ differs, as the simpler integrals appearing in Eq.~(\ref{intBoltz}) can be expressed in terms of Bessel functions of the second kind $K_n(z)$.
We note that compared with the standard perfect-fluid expressions,
\begin{equation}\label{eq:no}
N_{\rm eq}^\mu = n u^\mu,
\end{equation}
\begin{equation}\label{eq:to}
T_{\rm eq}^{\mu \nu} = \varepsilon u^\mu u^\nu - P \Delta^{\mu \nu},
\end{equation}
and the ``phenomenological'' form of the spin tensor \cite{Weyssenhoff:1947iua},
\begin{equation}\label{eq:so}
S_{\rm eq}^{\lambda, \mu \nu} = u^\lambda S^{\mu \nu},
\end{equation}
there are additional terms. Although seemingly their mathematical structure is typical of dissipative corrections, in this framework they appear already in the description of the perfect fluid; the entropy current, defined here as
\begin{equation}\label{1}
S_{\rm eq}^\mu =  T_{\rm eq}^{\mu \alpha} \beta_\alpha-\f{1}{2} \omega_{\alpha\beta} S_{\rm eq}^{\mu, \alpha \beta} -\xi N_{\rm eq}^\mu + {\cal N}^\mu, \quad {\cal N}^\mu = \coth\xi\,\,N^\mu,
\end{equation}
remains conserved. Our complete results for the Fermi-Dirac and Boltzmann tensor coefficients have been collected in Tables~\ref{tab:N}, \ref{tab:T}, and~\ref{tab:S} for the baryon current, the energy-momentum tensor, and the spin tensor, respectively.

Fig.~\ref{fig:coeff} shows a comparison between the Fermi-Dirac and Boltzmann cases. As can be readily seen in the logarithmic graphs, for constant particle mass $m$ and temperature $T$, the relative difference increases exponentially with the baryon chemical potential $\mu$ for most of its plotted range.
In the low $\mu$ regime, where the Fermi-Dirac distribution can be satisfactorily approximated by the Boltzmann distribution, the difference between the coefficient values in the two cases is negligible, but in the high $\mu$ regime it is significant. The relative differences between the two cases are largest for $n_2^\omega$ and $\epsilon_2^\omega$, reaching the order of $1$ for $\xi = \mu/T \approx 10$ (for the sample values of $m = 1000$ MeV and $T = 100$ MeV chosen).

\subsection{Tensors of dissipative spin hydrodynamics}

Following the Israel-Stewart method, we write down general nonequilibrium expressions as a sum of the equilibrium terms and dissipative corrections
\begin{equation}\label{2}
N^\mu = N_{\rm eq}^\mu + \delta N^\mu,\quad T^{\mu \nu} = T_{\rm eq}^{\mu \nu} + \delta T^{\mu \nu}, \quad S^{\mu, \alpha \beta} = S_{\rm eq}^{\mu, \alpha \beta} + \delta S^{\mu, \alpha \beta}.
\end{equation}
It is through dissipative terms that the spin-orbit interaction is introduced: $T^{\mu \nu}$ can now contain nonsymmetric parts, and the spin tensor is no longer seperately conserved
\begin{equation}
\p_\mu J^{\mu, \alpha \beta} = 0, \quad
J^{\mu, \alpha \beta} = x^\alpha T^{\mu \beta} - x^\beta T^{\mu \alpha} + S^{\mu, \alpha \beta},
\end{equation}
\begin{equation}\label{3}
\p_\mu S^{\mu, \alpha \beta} = T^{\beta \alpha} - T^{\alpha \beta}, \quad \p_\mu N^\mu = 0, \quad \p_\mu T^{\mu \nu} = 0.
\end{equation}
From Eqs.~(\ref{1}), (\ref{2}), (\ref{3}), we get
\begin{equation}
\p_\mu S^\mu = \delta N^\mu \p_\mu \xi + \delta T_s^{\mu \nu} \p_\mu \beta_\nu + \delta T_a^{\mu \nu} (\p_\mu \beta_\nu - \omega_{\nu \mu}) - \f{1}{2} \delta S^{\mu, \alpha \beta} \p_\mu \omega_{\alpha \beta};
\end{equation}
compare, e.g., \cite{Hattori:2019lfp, Becattini:2023ouz}.
To find the form of the deviations $\delta N^\mu$, $\delta T^{\mu \nu}$, $\delta S^{\mu, \alpha \beta}$, we use a~decomposition of general tensors of the given symmetry via projections along $u^\mu$ and separation into symmetric and antisymmetric parts, obtaining, through the Landau matching conditions (see \cite{Drogosz2024hybrid} for details),
\begin{align}\begin{split}\label{dev}
\delta N^\mu  &= V^\mu, \\
\delta T^{\mu\nu}_s &= -\Pi \Delta^{\mu\nu} + W^\mu u^\nu + W^\nu u^\mu + \pi^{\mu\nu},\\
\delta T^{\mu\nu}_a &= d^\mu_a u^\nu - d^\nu_a u^\mu + e^{\mu\nu}_a, \\
\delta S^{\lambda, \mu\nu} &= \Sigma^{\lambda\mu} u^\nu - \Sigma^{\lambda\nu} u^\mu + \phi^{\lambda \mu\nu},
\end{split}\end{align}
with 
\begin{align}\begin{split}
V^\mu &= b^\mu - n_t t^\mu,\quad \Pi = e - (P_0 + P_2^\omega + P_2^k) + (1/3)  P_{k\omega} (k^2 + \omega^2),\\ W^\mu &= d^\mu_s - P_t t^\mu, \quad
\pi^{\mu\nu} = e^{\LAB \mu\nu \RAB}_s - P_{k\omega} (\, k^{\LAB \mu} k^{\nu \RAB}  +\,\omega^{\LAB \mu} \omega^{\nu \RAB} ),\\ 
\Sigma^{\lambda \mu} &= i^{\lambda \mu} - \f{A}{2} t^{\lambda \mu}, \quad \phi^{\lambda\mu\nu} = j^{\lambda\mu\nu} - \f{A}{2} ( \Delta^{\lambda \mu} k^\nu - \Delta^{\lambda \nu} k^\mu).
\end{split}\end{align}
Angular brackets around a pair of indices denote the orthogonal, symmetric, and traceless part of the tensor, defined via a contraction with the projector $\Delta^{\mu\nu}_{\alpha\beta}$, $A^{\LAB \alpha \beta \RAB} \equiv \Delta^{\mu\nu}_{\alpha\beta} A^{\alpha \beta}  \equiv \f{1}{2} \LB
\Delta^{\mu}_{\alpha} \Delta^{\nu}_{\beta} +
\Delta^{\nu}_{\alpha} \Delta^{\mu}_{\beta} - \f{2}{3} 
\Delta^{\mu\nu} \Delta_{\alpha\beta} \RB A^{\alpha \beta}$. Squared brackets denote the antisymmetric part, $A^{[\mu\nu]} \equiv  \frac{1}{2} (A^{\mu\nu}-A^{\nu\mu})$. Angular brackets around a single index denote an orthogonal projection $T^{\alpha \beta \LAB \gamma \RAB \delta ...} \equiv \Delta^\gamma_\rho T^{\alpha \beta \rho \delta ...}$. Differential operators take precedence over the orthogonal projection. The nabla symbol denotes the transverse gradient $\nabla^\mu \equiv \Delta^{\mu \nu} \p_\nu$.

Equations (\ref{dev}) have a form that was analyzed in Ref.~\cite{Biswas:2023qsw}, and the coefficients $b^\mu$, $d_s^\mu$, $d_a^\mu$ $e$, $e^{\LAB \mu\nu \RAB}_s$, $e^{\mu \nu}_a$, $i^{\lambda \mu}$, $j^{\lambda \mu \nu}$ can be expressed in terms of gradients of hydrodynamic variables multiplied by kinetic coefficients, including the shear and bulk viscosity $\eta, \zeta$, the thermal conductivity $\kappa$, the coefficients $\lambda_a$ and $\gamma$ from  Ref.~\cite{Hattori:2019lfp}, and the coefficients $\chi_1 ,\chi_2, \chi_3, \chi_4$ introduced in Ref.~\cite{Biswas:2023qsw}.
To the first order in gradients, we find
\begin{align}\begin{split}\label{terms}
b^\mu &= \lambda \nabla^\mu \xi + n_t t^\mu, \quad
d^\mu_s = -\kappa (D u^\mu - \beta \nabla^\mu T ) + P_t t^\mu, \\
d^\mu_a &=\lambda_a \beta^{-1} 
(\beta D u^\mu + \beta^2  \nabla^\mu T - 2 k^\mu), \quad
e = {\bar P} -\zeta \theta - (1/3)  
P_{k\omega} (k^2 + \omega^2), \\
e^{\LAB \mu\nu \RAB}_s &= 2 \eta \sigma^{\mu\nu} + P_{k\omega} (\, k^{\LAB \mu} k^{\nu \RAB}  +\,\omega^{\LAB \mu} \omega^{\nu \RAB} ), \quad
e^{\mu \nu}_a =  \gamma 
\beta \nabla^{[ \mu} u^{\nu ]}, \\
i^{\lambda\mu} &= -\chi_1 \Delta^{\lambda\mu} u^\beta \nabla^\alpha \omega_{\alpha\beta}
- \chi_2 u_\nu \nabla^{\LAB \lambda} \omega^{\mu \RAB \nu} 
 -\chi_3 u_\nu \Delta^{ [ \mu}_\rho \nabla^{ \lambda ] } \omega^{\rho \nu} 
+\f{A}{2} t^{\lambda \mu}, \\
j^{\lambda\mu\nu} &=
\f{\chi_4 }{2} \nabla^{ \lambda} \omega^{\LAB \mu \RAB \LAB \nu \RAB} 
+\f{A}{2} ( \Delta^{\lambda \mu} k^\nu - \Delta^{\lambda \nu} k^\mu).
\end{split}\end{align}
Importantly, the spin polarization tensor $\omega_{\mu \nu}$, which in natural units is a dimensionless quantity, is not considered to be a gradient term. Already, the perfect-fluid case is based on an expansion in $\omega_{\mu \nu}$ to the second order. Gradient terms are only introduced when considering dissipation, and the order in gradients is counted separately to the order in $\omega_{\mu \nu}$. Therefore, in the expansion above, which is linear in gradients, terms such as $\nabla^\alpha \omega_{\alpha \beta}$ survive and are not neglected as higher-order contributions.

\section{Discussion}

The hybrid framework of spin hydrodynamics combines the perfect-fluid results of kinetic theory for particles with spin $\onehalf$ with the Israel-Stewart approach for including nonequilibrium processes. It involves a 
two-fold expansion: in the spin polarization tensor $\omega_{\mu \nu}$ (already in local equilibrium) and in gradients of hydrodynamic variables (when considering close-to-equilibrium dynamics).
If the spin polarization is nonvanishing, the perfect-fluid description contains seemingly dissipative, transverse terms.
However, genuine dissipative terms appear only at the level of the dissipative fluid. They are determined by the condition of positive entropy production.

This contribution focuses on summarizing the derivation of the tensors appearing in the hybrid framework, with the emphasis on Fermi-Dirac statistics. Thus, it complements the outline given in Ref.~\cite{Florkowski2025spinh}, which discusses in more detail the motivation behind the framework, as well as the modification of the usual thermodynamic relations of relativistic hydrodynamics that makes them consistent with the fact that the spin tensor (\ref{spintensor}) derived from kinetic theory, in contrast with the ``phenomenological'' spin tensor (\ref{eq:so}), contains parts orthogonal to the flow vector $u^\mu$.

We have shown that the difference between the tensor coefficients derived for the Fermi-Dirac statistics and their Boltzmann approximation is significant in the regime of high baryon chemical potential.

The coefficients derived herein can be employed in lieu of the Boltzmann approximation in future computer simulations, thereby enhancing their realism.
Although the functions $J_{mn}$, $\dotj_{mn}$ and $\ddj_{mn}$, unlike Bessel functions, are not a part of standard mathematical packages, their suitable tabulation and interpolation for use in practical codes presents very little technical difficulty.

The first numerical test of the hybrid framework was performed in a simple boost-invariant perfect-fluid case in Ref.~\cite{Drogosz2025boostinv}. That analysis can be extended to include the Fermi-Dirac statistics and dissipative terms. The framework should also be straightforward to implement
in numerical simulations of spin hydrodynamics in more realistic expansion models, similar to those presented in Refs.~\cite{Sushant2025prc, Sapna:2025yss}.

\section*{Acknowledgments}

\noindent
I thank my collaborators Wojciech Florkowski and Mykhailo Hontarenko for clarifying discussions. 


%


\end{document}